\documentclass[preprint]{aastex}
\usepackage{subfigure}
\pdfoutput=1

\slugcomment{}

\shorttitle{VLBA Astrometry of Cassini}
\shortauthors{Jones et al.}

\begin{document}

\title{VLBA Astrometric Observations of the \\
    Cassini Spacecraft at Saturn}

\author{Dayton L.~Jones}
\affil{Jet Propulsion Laboratory, California Institute of Technology, 
Pasadena, CA 91109}
\email{dayton.jones@jpl.nasa.gov}

\author{Ed Fomalont}
\affil{National Radio Astronomy Observatory, Charlottesville, VA
22903}

\author{Vivek Dhawan}
\affil{National Radio Astronomy Observatory, Socorro, NM 87801}

\author{Jon Romney}
\affil{National Radio Astronomy Observatory, Socorro, NM 87801}

\author{William M.~Folkner}
\affil{Jet Propulsion Laboratory, California Institute of Technology, 
Pasadena, CA 91109}

\author{Gabor Lanyi}
\affil{Jet Propulsion Laboratory, California Institute of Technology, 
Pasadena,CA 91109}

\author{James Border}
\affil{Jet Propulsion Laboratory, California Institute of Technology,
Pasadena, CA 91109}

\and

\author{Bob Jacobson}
\affil{Jet Propulsion Laboratory, California Institute of Technology, 
Pasadena, CA 91109}

\begin{abstract}
The planetary ephemeris is an essential tool for interplanetary spacecraft navigation, studies of solar system dynamics (including, for example, barycenter corrections for pulsar timing ephemeredes), the prediction of occultations, and tests of general relativity.  We are carrying out a series of astrometric VLBI observations of the Cassini spacecraft currently in orbit around Saturn, using the Very Long Baseline Array (VLBA).  These observations provide positions for the center of mass of Saturn in the International Celestial Reference Frame (ICRF) with accuracies $\sim$0.3 milli-arcsecond (1.5 nrad), or about 2 km at the average distance of Saturn.  This paper 
reports results from eight observing epochs between 2006 October and 2009 April.  These data are combined with two VLBA observations by other investigators in 2004 and a Cassini-based gravitational deflection measurement by Fomalont et al. in 2009 to constrain a new ephemeris (DE 422).  The DE 422  post-fit residuals for Saturn with respect to the VLBA data are generally 0.2 mas, but additional observations are needed to improve the positions of all of our phase reference sources to this level.  Over time we expect to be able to improve the accuracy of all three coordinates in the Saturn ephemeris (latitude, longitude, and range) by a factor of at least three.  This will represent a significant improvement not just in the Saturn ephemeris but also in the link between the inner and outer solar system ephemeredes and in the link to the inertial ICRF. 
\end{abstract}


\section{Introduction}

Planetary ephemerides are used for multiple purposes including dynamical mass determination for solar system bodies, pulsar timing, high-precision tests of general relativity, and inter-planetary spacecraft navigation.  During the past several decades a series of increasingly accurate ephemerides have been developed at the Jet Propulsion Laboratory (JPL) by adding new data types such as Very Large Array (VLA) astrometry (\citet{Muhleman85}; \citet{Muhleman86}), spacecraft tracking \citep{Duxbury89}, and radar range measurements \citep{Campbell78} to historical and modern optical observations. 

Accurate ephemerides are one of the basic tools of observational astronomy, in the same sense as star catalogs and redshift surveys.  They represent a community resource whose value is proportional to their accuracy, and whose accuracy requires regular observational support to maintain and improve.  A specific example is the great improvement between timing distances and kinematic distances for pulsars when using the newer DE405 ephemeris \citep{Standish04} compared with the older DE200 ephemeris ({\it e.g.}, \citet{Verbiest08}).

The orbits of the inner planets are very accurately tied together with the current data set.  For example, the angular ephemeris errors for Mars with respect to Earth are typically 0.2 milli-arcsecond (mas) or 1 nrad \citep{IAU09}.  However, the outer planets are not as well tied to the inner planets (or each other) because there have been fewer opportunities to supplement optical observations with high precision spacecraft radio tracking data.  
The Pioneer and Voyager missions provided essentially single data points during their flybys of the outer planets, and the Galileo mission to Jupiter was severely constrained by the loss of its high gain antenna.  This restricted Galileo downlink signals to a relatively low frequency (2.3 GHz) and a low signal/noise ratio.  As a result, VLBI observations of Galileo had accuracies of only $\sim$5-10 mas \citep{Jacobson99}.  
Thus, the Cassini mission to Saturn (http://saturn.jpl.nasa.gov/index.cfm) is our first opportunity to incorporate high-accuracy data from a spacecraft orbiting an outer planet for an extended period.  For reference, an angle of 0.1 mas (0.5 nrad) corresponds to about 750 meters at the average distance of Saturn from Earth. 

Our goal is to improve the position of Saturn in the International Celestial Reference Frame (ICRF, see \citet{Ma98}) through phase-referenced VLBI observations of Cassini using the Very Long Baseline Array (VLBA)\footnote{The VLBA is operated by the National Radio Astronomy 
Observatory (NRAO).} at 8.4 GHz (X band) combined with Cassini orbit determinations.  The Cassini orbit can be determined to about 2 km at apoapse and 0.1 km at periapse relative to the center of mass of Saturn with range and Doppler tracking by the Deep Space Network \citep{Antreasian08}.  The future Juno mission to Jupiter should allow a similar application of phase-referenced VLBI to improve the Jupiter ephemeris.  Combined with our new data for Saturn this will lead to a better model for the gravitational interactions, and the orbital evolution, among all the outer planets.  A covariance analysis \citep{Standish06} at JPL shows that with only a few years of VLBA data, the ephemeris improvement for Saturn extends for decades.  

\section{Observations}

\begin{figure}
\includegraphics[angle=-90,scale=0.60]{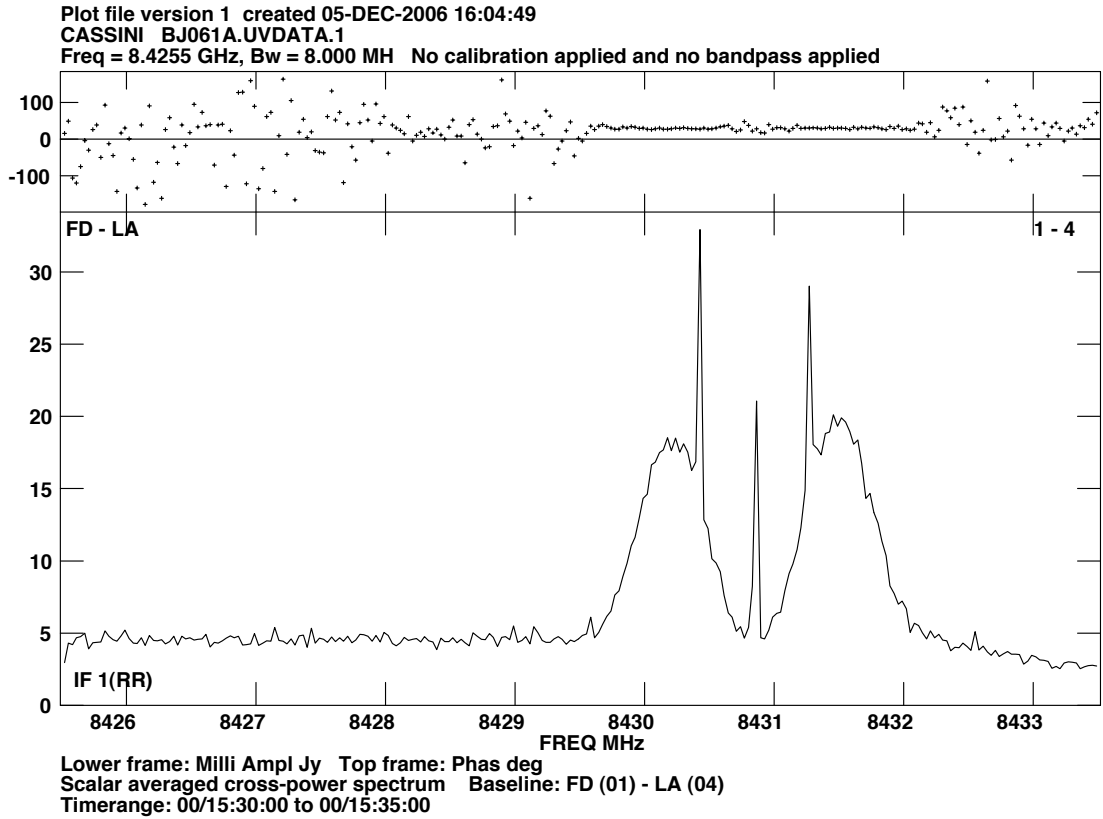}
\caption{Cassini fringe phase (top) and amplitude (bottom) from a single baseline for 
a typical 1-2 minute scan during our first epoch.  The narrow amplitude peaks are the carrier and subcarriers 
of the spacecraft signal, and the the broad peaks are created by telemetry modulation.  The total width of the 
signal varies with data rate.  During this epoch is was approximately 2.5 MHz.  \label{fig1}}
\end{figure}

We have observed Cassini with the VLBA at eight epochs, each typically four hours long, during the past three years.  Table \ref{tab1} lists the dates of each 
epoch and the VLBA antennas that were used.  The VLBA consists of ten radio antennas, each 25 meters in diameter, located at sites from the US Virgin Islands to Hawaii.  It has demonstrated an unrivaled astrometric precision of less than 10 micro-arcseconds ($\mu$as) in favorable circumstances (e.g., \citet{Fomalont03}).  The spacecraft's 8 GHz signals provide more than adequate SNR as shown in Figure \ref{fig1}. 

\clearpage
    
\begin{deluxetable}{ccc}
\tablewidth{0pt}
\tablecaption{Observing Epochs and VLBA Antennas Used \label{tab1}}
\tablehead{
\colhead{Epoch} & \colhead{Obs.~Date} & \colhead{VLBA Antennas}}    
\startdata    
BJ061A & 2006 Oct 11 & SC, HN, NL, FD, LA, PT, KP, OV, MK \\  
BJ061B & 2007 Mar 1 & SC, HN, NL, FD, LA, PT, KP, OV, BR, MK \\  
BJ061C & 2007 Jun 7 & SC, HN, NL, FD, LA, PT, KP, OV, BR, MK \\
BJ061D & 2008 Jan 12 & SC, HN, NL, LA, PT, KP, OV, BR, MK \\
BJ061E & 2008 Jun 13 & SC, HN, FD, LA, KP, OV, MK \\  
BJ061F & 2008 Aug 1 & SC, HN, NL, FD, LA, PT, KP, OV, BR, MK \\
BJ061G & 2008 Nov 11 & HN, NL, FD, LA, PT, KP, OV, BR, MK \\
BJ061H & 2009 Apr 24 & SC, HN, NL, FD, LA, PT, KP, OV, BR, MK \\
\enddata
\tablecomments{The VLBA antenna locations are:  SC = St.~Croix, US Virgin Islands; 
HN = Hancock, NH; NL = North Liberty, IA; FD = Fort Davis, TX; LA = Los Alamos, NM; 
PT = Pie Town, NM; KP = Kitt Peak, AZ; OV = Owens Valley, CA; BR = Brewster, WA; 
MK = Mauna Kea, HI. }
\end{deluxetable}

\clearpage

We used standard phase-referencing techniques (\citet{Lestrade90}; \citet{Guirado97};  \citet{Guirado01}; \citet{Fomalont06}) with rapidly alternating scans between Cassini and angularly nearby reference sources (see Table \ref{tab2}).  
In addition, we observed several strong sources spread over the sky during each epoch to allow better calibration of the troposphere (\citet{Lestrade04}; \citet{Mioduszewski04}).  We did not attempt to improve the phase calibration by employing multiple phase reference sources when possible, as has been demonstrated by \citet{Fomalont05}, but this is an option for future observations.  

\begin{figure}
\includegraphics[angle=0,scale=0.70]{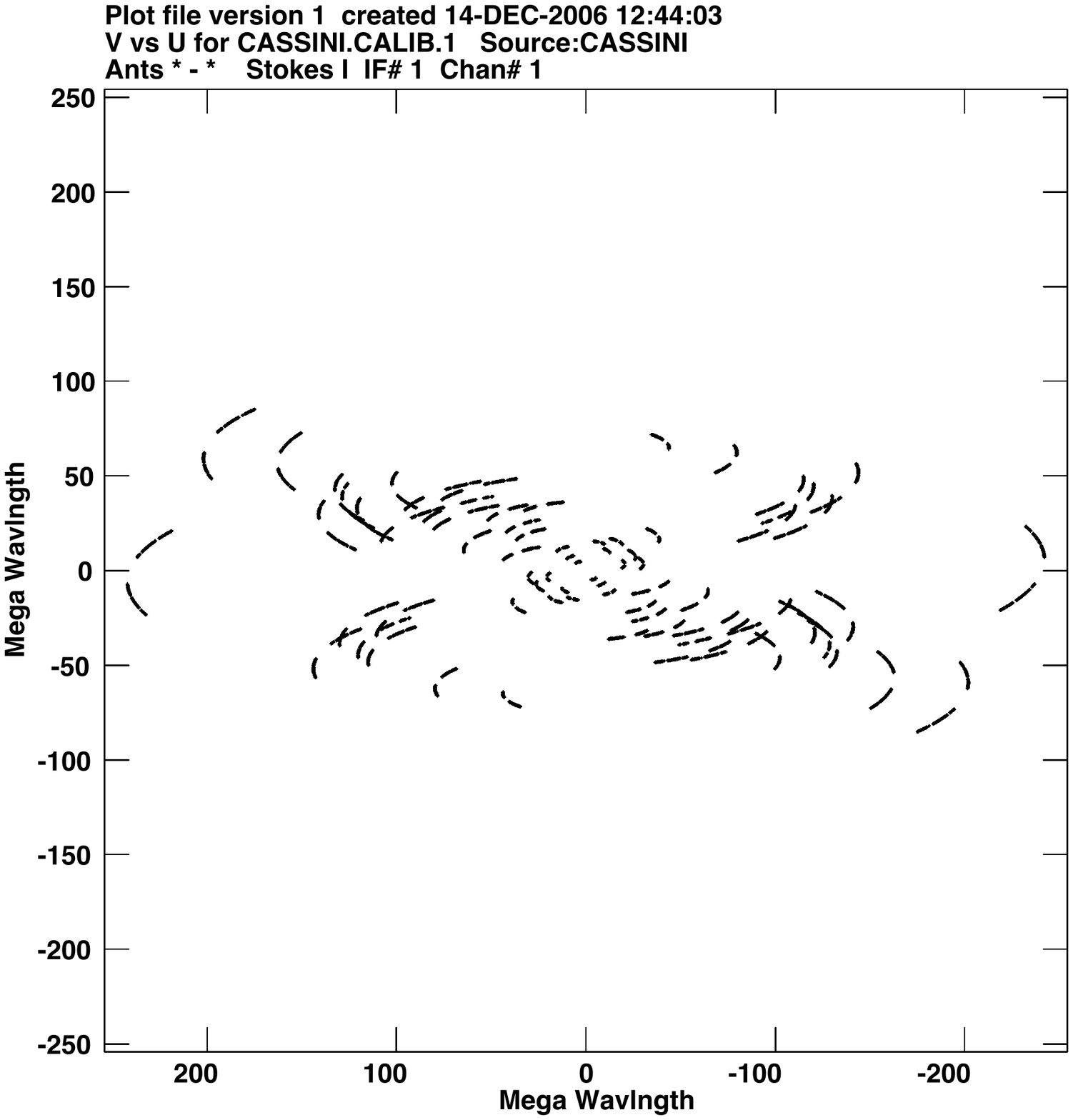}
\caption{(U,V) sampling for Cassini during our first epoch.  The sampling 
for the phase reference source is nearly identical. \label{fig2}}
\end{figure}

Figure \ref{fig2} shows the projected baselines for one of our experiments (BJ061A).  For some epochs, including this one, the most northerly VLBA antenna site, in Brewster, WA, was not able to participate.  This caused a reduction in north-south resolution.  Nevertheless, the over-all baseline coverage is good for a low-declination radio source.  It is sufficient for unambiguous phase referencing.  

The accuracy of the ICRF is approximately 0.02 mas, based on $\sim$200 defining sources each with a typical position error of 0.1-0.2 mas.  Individual sources may have larger errors at some epochs due to changes in source structure (e.g., \citet{Fey04}; \citet{Porcas09}), but the imaging capability of the VLBA allows us to detect and correct for such changes.  To maximize the accuracy of our link between the Saturn ephemeris and the ICRF we have devoted a separate observing epoch to tying the phase reference sources to 10-20 high-quality ICRF sources.  (The results from this epoch will be reported separately.)  This reduced our sensitivity to the structure of individual sources.  We expect, based on previous experience with phase referenced VLBI (e.g., \citet{Lestrade99}), that an absolute accuracy of 0.1 mas with respect to the ICRF can be achieved for our calibration sources.  Thus, the final combined error of our Cassini positions with respect to the ICRF will be approximately 0.2 mas, consistent with the level of VLBI spacecraft tracking error predicted by \citet{Lanyi07}.  Because the linking of our phase references sources to the grid of ICRF sources is not complete, the current error for Cassini positions in the ICRF is estimated to be 0.3 mas or greater, depending on the specific phase reference source used.  One of our phase reference sources has an ICRF position error less than 0.1 mas, while some of the other reference sources have ICRF positions errors several times larger.  Future observations to maintain and improve the ICRF catalog will continue to reduce the position errors of these sources.

Calibration, editing, and other data reduction tasks were carried out using the Astronomical Image Processing System (AIPS)\footnote{AIPS is provided and supported by the National Radio Astronomy Observatory.}.  The only exception was the production of total delays for use by JPL software.  This required a specific file format, which was created by combining information from AIPS output tables with a program written by E. Fomalont.

The analysis of data from a phase-referenced VLBI experiment involves multiple steps because there are many types of error that must be calibrated and removed from the data (e.g., \citet{Lanyi05}).  The following subsections describe these analysis steps.

\section{Experiment Scheduling}

There are a number of constraints that must be satisfied when scheduling each epoch. 
These are:
\begin{enumerate}
\item We need to avoid periods near Saturn conjunction with the Sun.
\item Cassini must be transmitting rate telemetry to the Deep Space 
Network station at Goldstone, CA, at X-band (8.4 GHz).
\item We need to avoid spacecraft trajectory correction maneuvers, moon flybys, 
and ring occultations.
\item We need a reasonably strong and compact phase reference source 
within about 2$^{\circ}$ of Cassini's position.
\item We need to observe multiple strong sources covering a wide range 
of elevation angles for troposphere calibration during each epoch.
\end{enumerate}

Because Saturn reverses its apparent direction of motion on the sky every year, 
it is possible with careful scheduling to use the same phase reference source 
during multiple epochs.  
This reduces the number of phase reference sources that need to be tied to the 
ICRF, and minimizes relative errors between epochs due to source structure 
or ICRF offset differences between difference reference sources. 

\clearpage
    
\begin{deluxetable}{ccccc}
\tablewidth{0pt}
\tablecaption{Observing Epochs and Phase Reference Sources \label{tab2}}
\tablehead{
\colhead{Epoch} & \colhead{Date} & \colhead{Reference Source} & 
\colhead{Angular Separation} & \colhead{Flux Density} \\
 & & & \colhead{(deg)} & \colhead{(Jy)}}    
\startdata    
BJ061A & 2006 Oct & J0931+1414 & 2.5 & 0.15 \\
BJ061B & 2007 Mar & J0931+1414 & $<$ 2 & 0.15 \\
BJ061C & 2007 Jun & J0931+1414 & $<$ 2 & 0.15 \\
BJ061D & 2008 Jan & J1025+1253 & 3.5 & 0.23 \\
BJ061E & 2008 Jun & J1025+1253 & $\sim$ 1 & 0.23 \\
BJ061F & 2008 Aug & J1025+1253 & $<$ 3 & 0.23 \\
BJ061G & 2008 Nov & J1127+0555 & $<$ 0.5 & 0.10 \\
BJ061H & 2009 Apr & J1112+0724 & $<$ 0.5 & 0.20 \\
\enddata
\end{deluxetable}

Note that we did not always have a phase reference source within $2^{\circ}$ of 
Cassini, as desired.  There is often a tradeoff between angular separation and 
reference source flux density, and we did not consider sources with flux densities 
less than 100 mJy to ensure an adequate signal/noise ratio.  In retrospect 
this flux density cutoff may have been too conservative. 

During each observing epoch we alternated scans between Cassini and 
the phase reference source every 2-3 minutes, and included occasional 
scans of a strong source for instrumental delay and bandpass calibration.  
In addition, $\sim 10$ strong sources covering as wide an elevation range 
as possible were observed during a 30-40 minute period to allow a zenith 
troposphere delay to be estimated for each of the VLBA antenna sites. 

The data from each epoch were stored prior to correlation until a 
reconstructed orbit solution for Cassini was available from JPL.  The 
Cassini orbit is determined from Doppler tracking by the Deep Space 
Network, and provides the {\it a priori} positions used for Cassini 
during correlation.  Because Cassini transmits in right circular polarization 
only, we used only the R-R corralator output in our analysis.  To avoid 
decorrelation over time or frequency, and to accurately measure phase 
across the relatively narrow bandwidth of the Cassini signal, we used 
1-second integrations and 256 spectral channels across each 8-MHz 
IF band.  

The four available IF bands were separated in frequency by 
up to 462 MHz to improve the multi-band delay response.  Small changes 
in the reference frequency were necessary at each epoch to keep the 
Cassini signal away from the edges of IF band 1.  (The frequency of 
the Cassini signal changes by a few MHz when switching between a 
one-way downlink and a two-way coherent link.)   The center 
frequencies for each IF band were approximately 8.428, 8.500, 8.790, 
and 8.890 GHz.

\section{Astrometric Position Deteminations}

\subsection{A Priori Calibration}

Initial delays were calculated from the geometric model used during correlation
\citep{Romney99}.
The {\it a priori} spacecraft position and proper motion at the start of our 
observations were corrected, if necessary, from the values used during 
correlation.  (The most accurate Cassini orbit solutions were not available 
until a few weeks after a particular date.)  A subtle but important part of 
the geometric model is the difference in general relativity corrections 
for signals propagating through gravitational fields in the solar system, 
particularly those of the Sun and Saturn. 
Unlike the phase reference sources, Cassini can not be assumed to be 
at infinity; only part of the solar system gravitational field applies to 
signals from Cassini.  

Amplitude calibration was based on recorded system temperatures 
along with previously determined antenna gain curves.  Additional 
amplitude corrections were applied to remove the effect of using two-bit 
quantization when recording signals at the VLBA antennas.  Phases were 
corrected for parallactic angle and for improvements in the values of 
Earth orientation 
parameters (UT1 and polar motion).  Instrumental delay offsets were 
calibrated using either recorded phase calibration tones or by fringe 
fitting a strong calibration source and applying the resulting delay 
corrections to all scans.  This aligned phases within the 
four 16-MHz-wide IF bands.  All data were examined for discrepant 
values or interference, and to verify that the 
{\it a priori} calibration had been applied correctly.  

\subsection{Ionosphere Delay Calibration}

Our observations did not include very widely separated frequency bands, 
so the dispersive ionosphere delay (and Faraday rotation) was calibrated 
using global maps of zenith total electron content (TEC) determined from GPS 
networks at two-hour intervals.  Linear interpolation between global TEC 
maps bracketing an observation was used, including a longitude correction 
to account for the fact that the ionosphere should be approximately fixed 
with respect to the Sun, not the Earth.

\subsection{Troposphere Delay Calibration}

Multi-band delays were calculated for a sample of $\sim 15$ strong sources 
observed in a rapid sequence over a wide range of elevation angles.  
A linear fit to the four IF phases was used to determine the multi-band 
delay for each source.  At least three IFs were required to have good data.    
Zenith troposphere delays and clock errors (offsets and rates) were fit to 
the measured multi-band delays 
using the \citet{Chao74} mapping function (see \citet{Sovers98}; \citet{Mioduszewski04}).  
This calibration removed the residual 
tropospheric delay, thus aligning the phases between all four IF bands.  

\begin{figure}
\plottwo{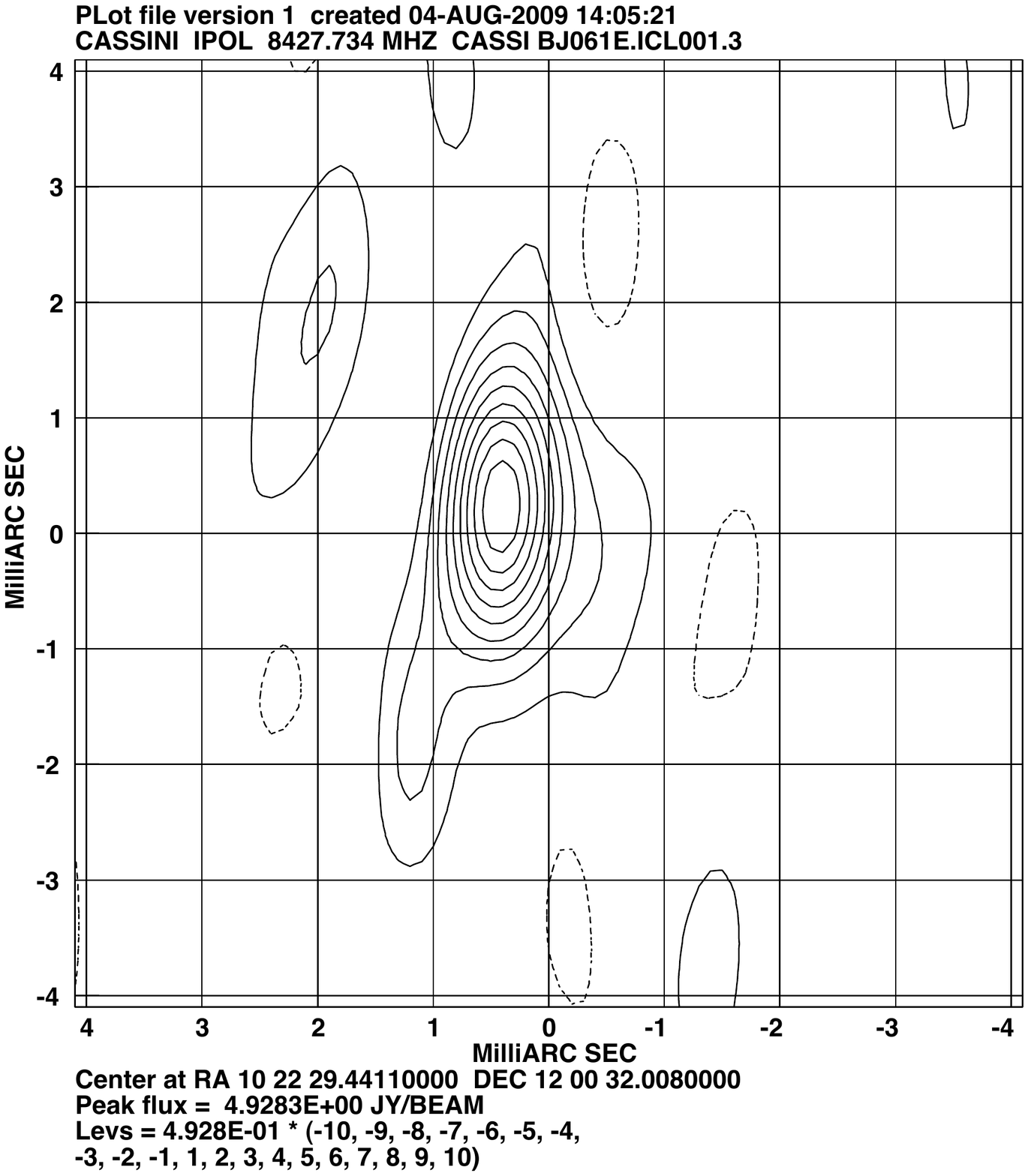}{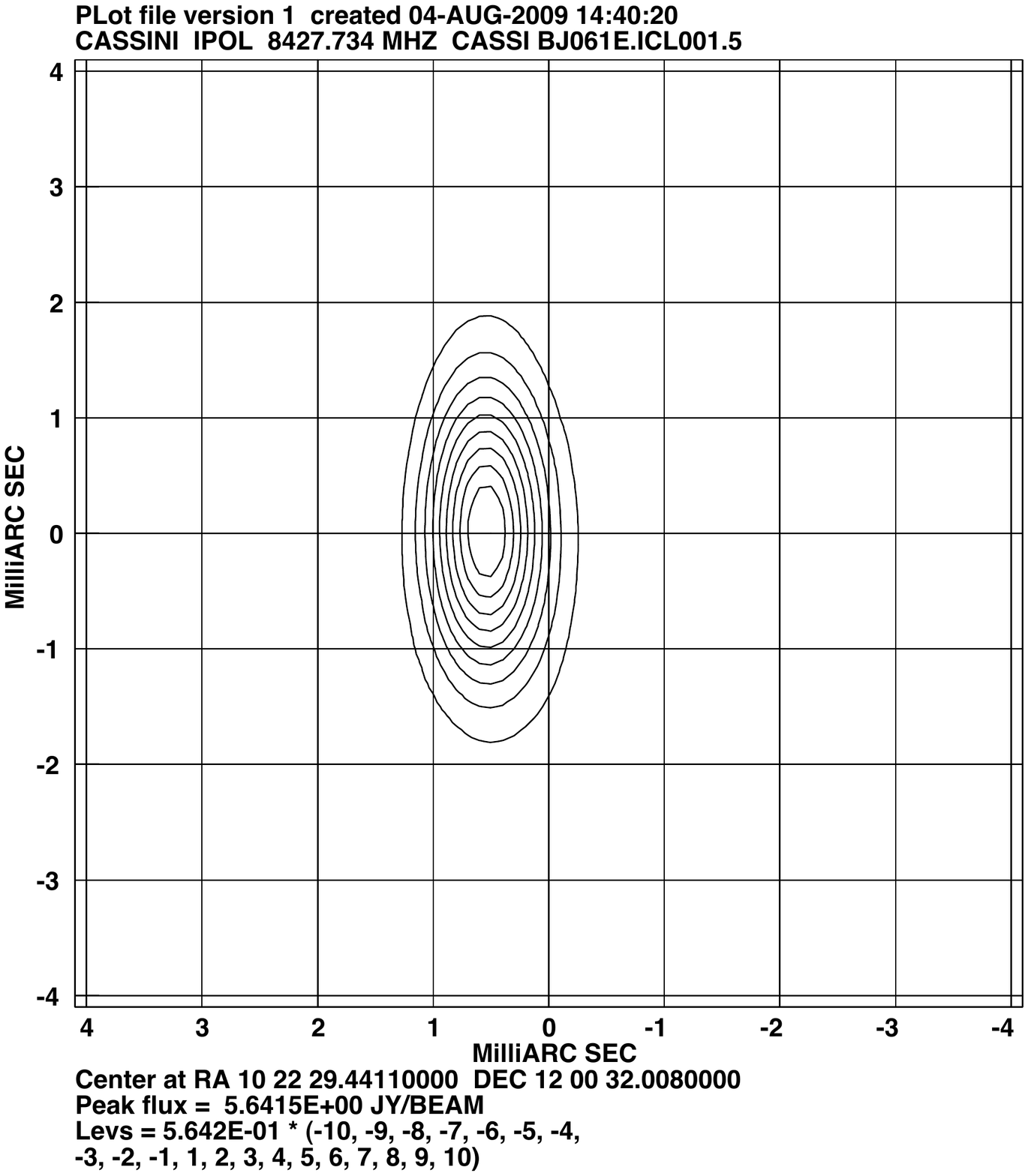}
\caption{Effect of manual troposphere delay calibration.  The left panel is 
the phase-referenced Cassini image from epoch 5 without troposphere 
calibration.  The right panel shows the improvement when troposphere delays 
were estimated for each antenna using calibrators observed over a wide range 
of elevation angles.  Note the reduction in scattered power and the higher 
peak flux density (4.93 Jy/beam on left, 5.64 Jy/beam on right).
Note also the small shift in declination between the two images. \label{fig3}}
\end{figure}

Figure \ref{fig3} shows the improvement in image quality produced by 
this manual troposphere delay calibration.  Note that the correction is 
mainly in the declination direction, as would be expected for a low declination 
source.  The position offset is about 0.2 mas, corresponding to a difference 
in differential delay of a few mm on the longer VLBA baselines.   
Seven of the eight epochs 
reported in this paper were improved in this way.  However, for one epoch 
(BJ061D) all attempts to calibrate troposphere delays manually produced 
poorer results, and we relied on the basic correlator troposphere model 
for this epoch.  

\subsection{Bandpass Calibration}

A strong calibration source was observed during each epoch to determine
any remaining antenna-based delays and to calibrate the bandpass response.  
After this step, scan-averaged phases were inspected for all calibration sources 
and baseline to verify that the phases were constant with frequency within and 
between IF bands.

\subsection{Phase-Referenced Imaging}

Self-calibration with a point source model was used to optimize the phases of the 
phase reference calibrator, and then these phases were 
applied to the Cassini visibilities.  No self-calibration was 
applied to the Cassini data.  
At this state images were 
made of both the 
phase reference source and Cassini.  The 
phase reference source, by self-calibration, was always 
located at it's {\it a priori} sky position.  We used the phase calibrator image 
to verify that a point source model was adequate.  

The location of 
the peak signal in the Cassini image was measured with the AIPS task 
maxfit and 
used to shift the image to place Cassini at (or very near) 
the phase center.  
Two-dimensional quadratic fits to the peaks in images of Cassini provide 
positions relative to the nominal phase center with formal errors under 
0.02 mas in both coordinates, far smaller than our systematic errors. 
The post-shift baseline phases were examined to verify that the removal 
of the position offset had centered the Cassini signal at the {\it a prioir} 
position.

\begin{figure}
  \begin{center}
    \subfigure[Epoch BJ061A]{\label{fig:egde-a}\includegraphics[scale=0.3]{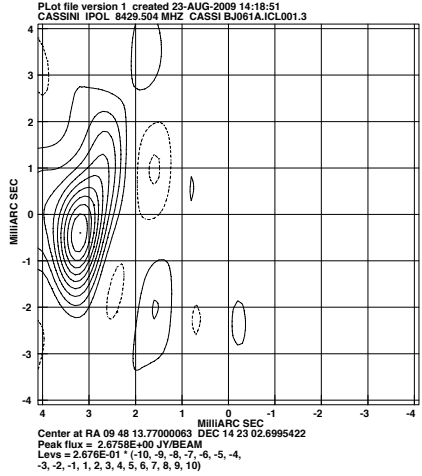}}
    \subfigure[Epoch BJ061B]{\label{fig:egde-b}\includegraphics[scale=0.3]{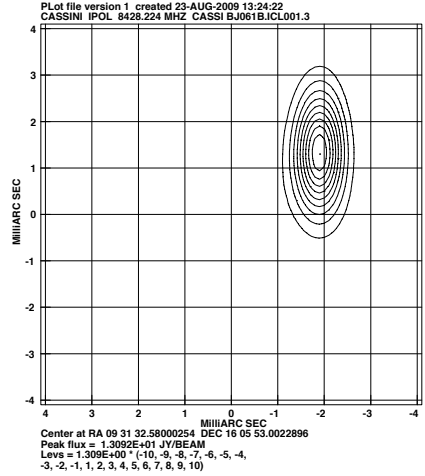}} \\
    \subfigure[Epoch BJ061C]{\label{fig:egde-c}\includegraphics[scale=0.3]{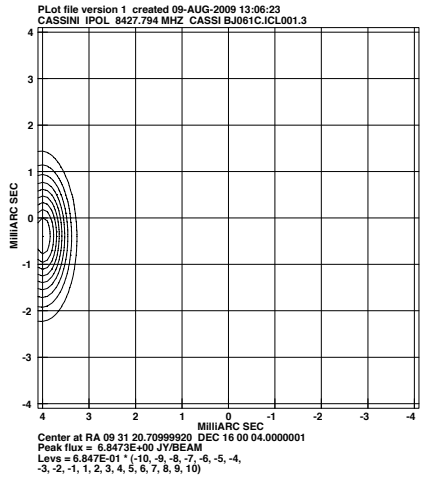}}
    \subfigure[Epoch BJ061D]{\label{fig:egde-d}\includegraphics[scale=0.3]{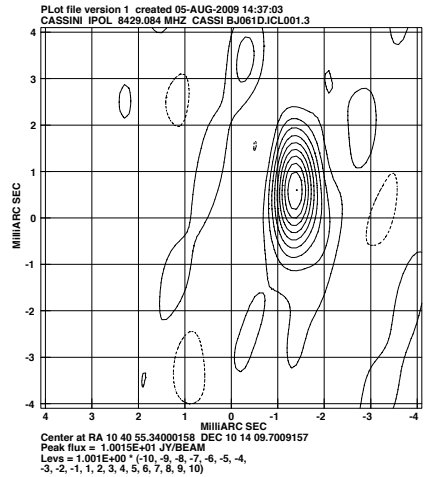}}
  \end{center}
\caption{Phase-referenced images of Cassini from the first 
four epochs.  The large and variable position offsets are 
caused by a one-second error in the spacecraft orbit 
calculations.  The peak flux densities are 2.68 Jy/beam (upper left), 
13.09 Jy/beam (upper right), 6.85 Jy/beam (lower left), and 
10.02 Jy/beam (lower right).  In all cases the contour levels are 
-20, -10, 10, 20, 30, 40, 50, 60, 70, 80, and 90\% of the peak flux density.  
\label{fig4}}
\end{figure}

\begin{figure}
  \begin{center}
    \subfigure[Epoch BJ061E]{\label{fig:egde-e}\includegraphics[scale=0.3]{BJ061ET_CASS_ICLN1.pdf}}
    \subfigure[Epoch BJ061F]{\label{fig:egde-f}\includegraphics[scale=0.3]{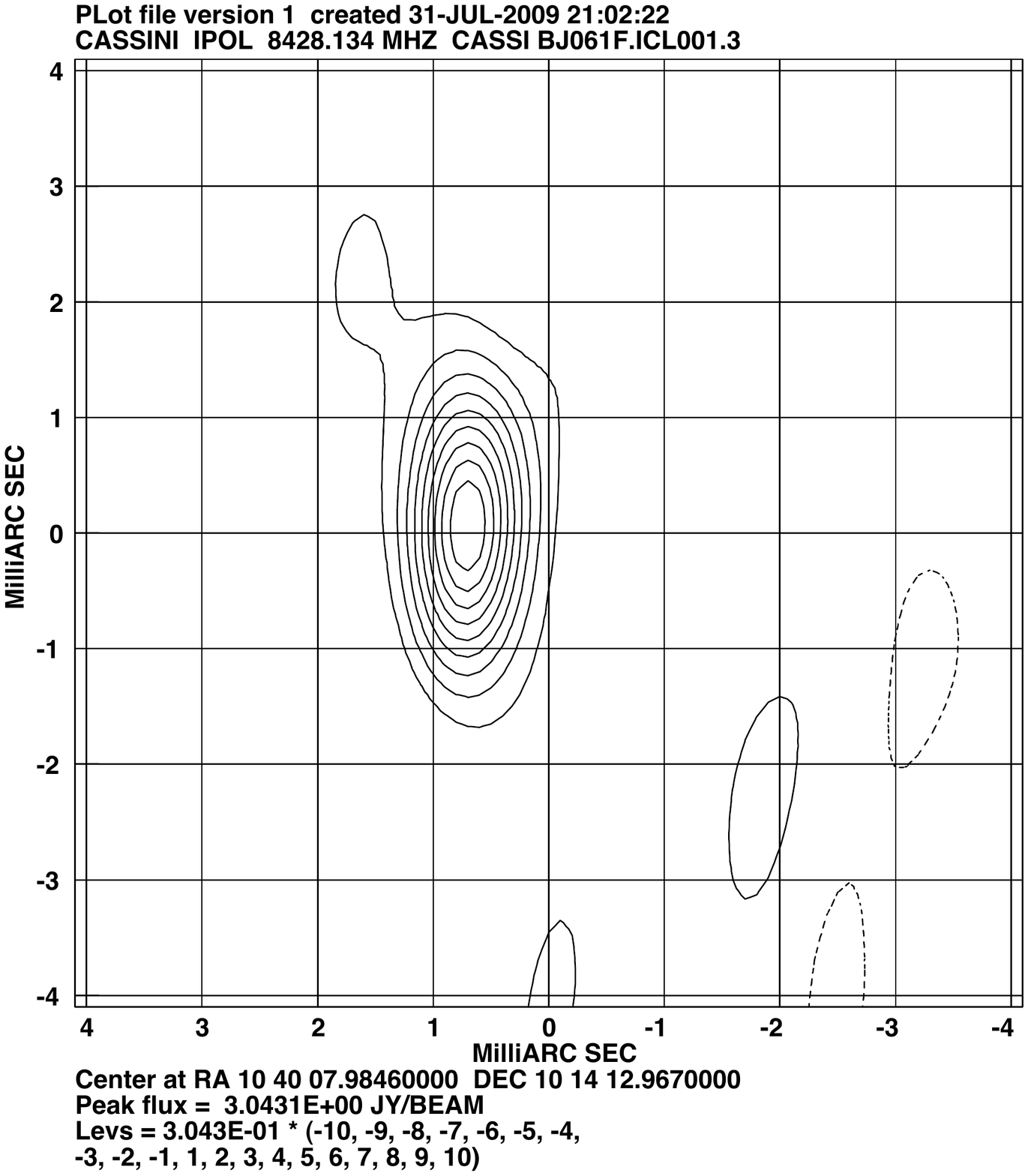}} \\
    \subfigure[Epoch BJ061G]{\label{fig:egde-g}\includegraphics[scale=0.3]{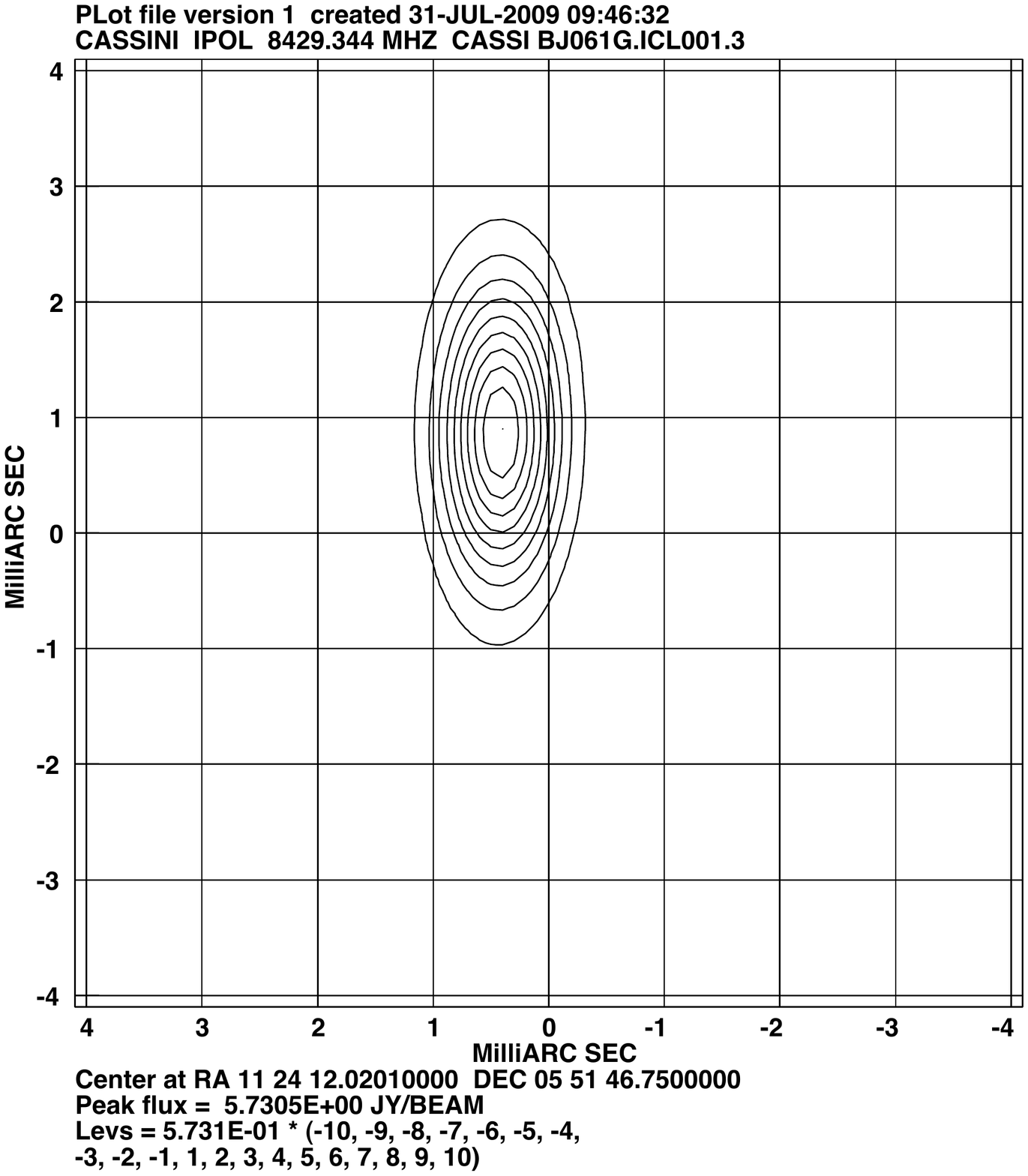}}
    \subfigure[Epoch BJ061H]{\label{fig:egde-h}\includegraphics[scale=0.3]{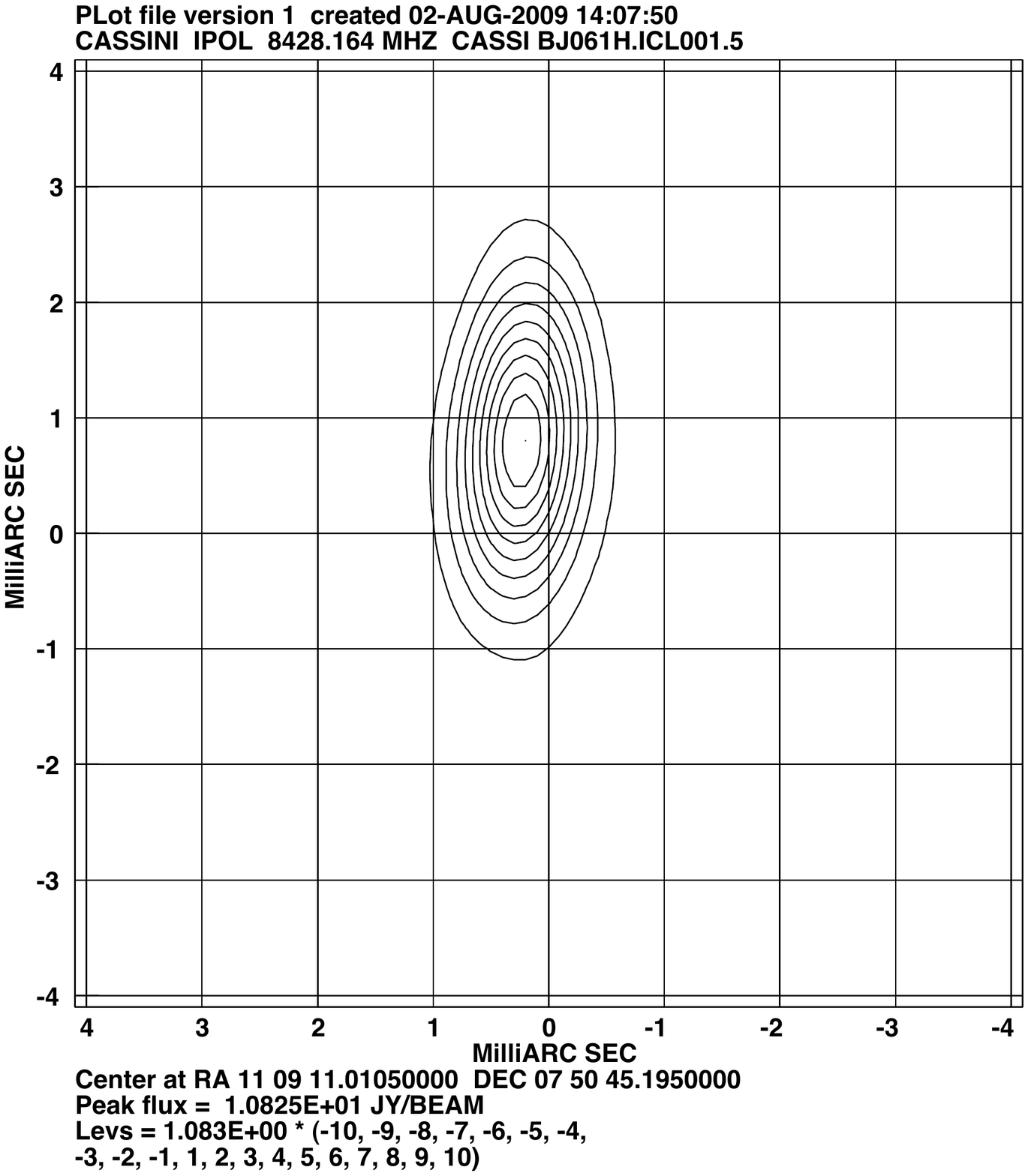}}
  \end{center}
\caption{Phase-referenced images of Cassini from the 
last four epochs.  The peak flux densities are 5.64 Jy/beam (upper left), 
3.04 Jy/beam (upper right), 5.73 Jy/beam (lower left), and 
10.83 Jy/beam (lower right).  In all cases the contour levels are 
-20, -10, 10, 20, 30, 40, 50, 60, 70, 80, and 90\% of the peak flux density.  
\label{fig5}}
\end{figure}

Figures \ref{fig4} and \ref{fig5} show the unshifted images of 
Cassini for all eight epochs.  The first four epochs, in Figure \ref{fig4}, 
all have large offsets from the phase center.  This was caused by 
a one-second error in the time used when calculating the spacecraft
position during correlation, the result of ignoring a leap second.  This 
error was discovered in 2008 March, but re-correlation with a corrected 
correlator model was not possible because the raw data recordings were no 
longer available.  The total 
delays were unaffected by this offset.  
The last four epochs, in Figure \ref{fig5}, show much smaller 
position offsets after the leap second error was found and corrected.  

\subsection{Phase Reference Source Positions}

The position of the peak 
Cassini signal was measured with respect to its image phase center, which in turn depended on the 
assumed position of the phase reference source and the geometric model.  Thus, any error in the 
phase reference source position produces a corresponding error in the position of Cassini.  We used 
the best available {\it a priori} positions for our phase reference sources during data analysis, but 
improved positions have recently become available as part of the ICRF2 catalog \citep{Fey09}.  

Table \ref{tab3} shows the ICRF2 position of our primary phase reference sources.  The maximum 
difference between our {\it a priori} reference source positions and the ICRF2 positions is 0.16 mas 
(for J0931+1414 in right ascension); all other position differences are less than 0.1 mas.  Note, however, 
that the ICRF2 position errors vary by more than an order of magnitude between sources.  Future 
observations will continue to improve the less accurate source positions.  

\clearpage
    
\begin{deluxetable}{ccc}
\tablewidth{0pt}
\tablecaption{Primary Phase Reference Sources \label{tab3}}
\tablehead{
\colhead{Source}  & \colhead{RA (J2000)}  &
\colhead{DEC (J2000)}}    
\startdata    
J0931+1414 & 09$^{\rm h}$31$^{\rm m}$05$^{\rm s}$.342445 $\pm0^{\rm s}.000029$  & 14$^{\circ}$14\arcmin16.51897\arcsec  $\pm$0.00090\arcsec \\
J1025+1253 & 10$^{\rm h}$25$^{\rm m}$56$^{\rm s}$.285370 $\pm0^{\rm s}.000004$  & 12$^{\circ}$53\arcmin49.02201\arcsec  $\pm$0.00008\arcsec \\
J1112+0724 & 11$^{\rm h}$12$^{\rm m}$09$^{\rm s}$.558539 $\pm0^{\rm s}.000016$  & 07$^{\circ}$24\arcmin49.11804\arcsec  $\pm$0.00050\arcsec \\
J1127+0555 & 11$^{\rm h}$27$^{\rm m}$36$^{\rm s}$.525564 $\pm0^{\rm s}.000066$  & 05$^{\circ}$55\arcmin32.05999\arcsec  $\pm$0.00172\arcsec \\
\enddata
\tablecomments{Positions and errors are from \citet{Fey09}.}
\end{deluxetable}

A recurring issue in radio astrometry is the positional stability of the radio cores 
used to define the ICRF and for narrow-angle phase referenced measurements. 
VLBI group delays give the position of the base of a radio jet, while VLBI phase delays 
give a frequency-dependent offset from the group delay position because of 
frequency-dependent opacity along the jet.  The ICRF is based on group delay 
measurements, and thus is relatively insensitive to jet opacity effects.  Our phase 
referenced astrometric measurements of Cassini, however, are directly affected 
by any variations in the apparent centroid position of our phase reference sources.

Phase and group delay positions are typically offset by 0.17 mas at 8.4 
GHz \citep{Porcas09}.  This offset is comparable to tropospheric calibration errors, 
and errors in the ICRF/ICRF2 positions of individual sources in many cases.  

\subsection{Total Delays}

The geometric model, including {\it a priori} antenna and 
source positions and the measured delays, plus the residual 
delay from the antenna phases determined from the shifted 
Cassini image, 
were output from AIPS in a series of tables.  These tables 
were text listings of the AN (antenna geometry), SU (source 
geometry), CL (correlator model and image position shift), 
and SN (residual phases) calibration tables associated 
with a visibility data file AIPS.  They were used 
to generate total delay values that could be used by 
JPL navigation software.  

\begin{figure}
\plottwo{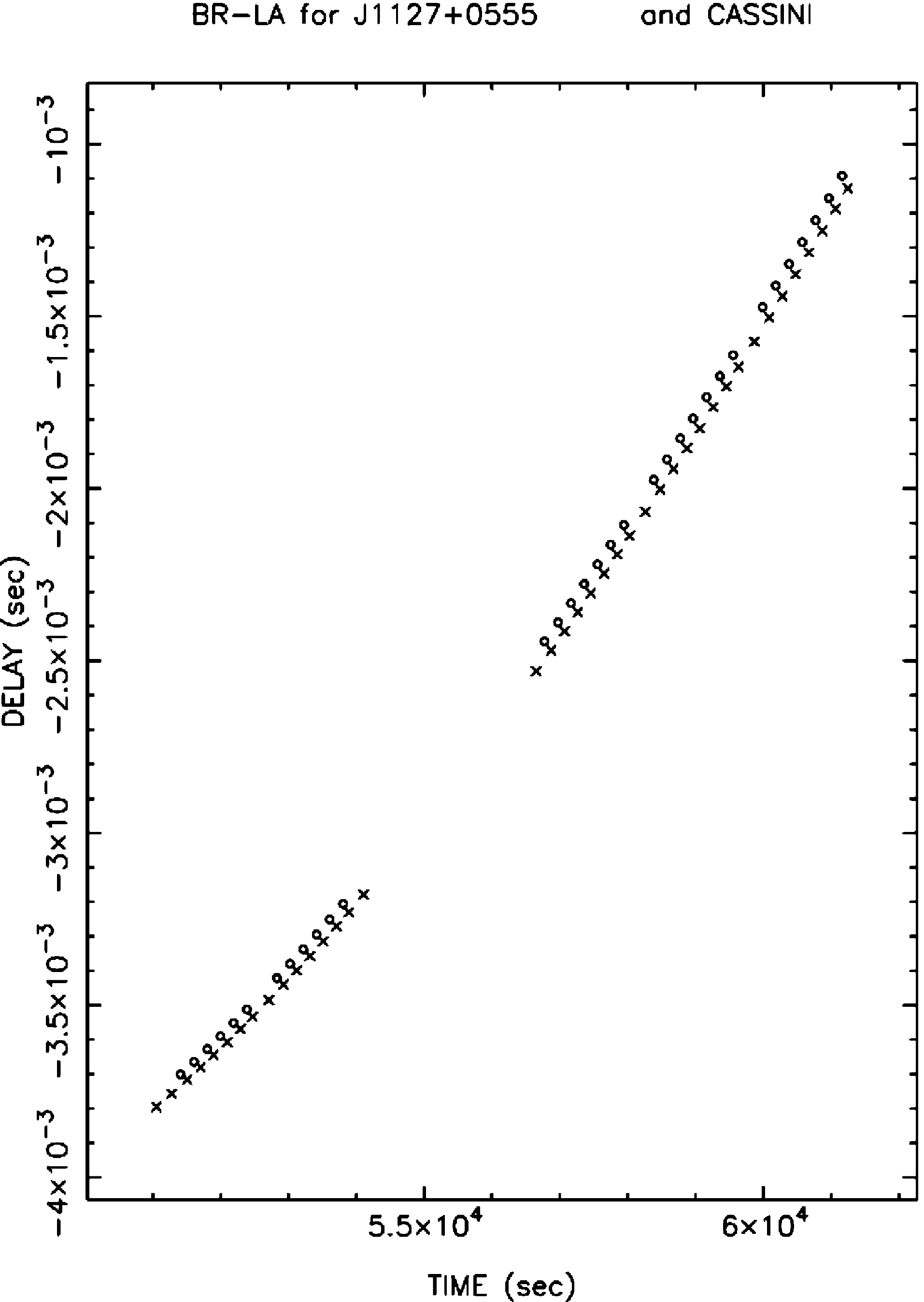}{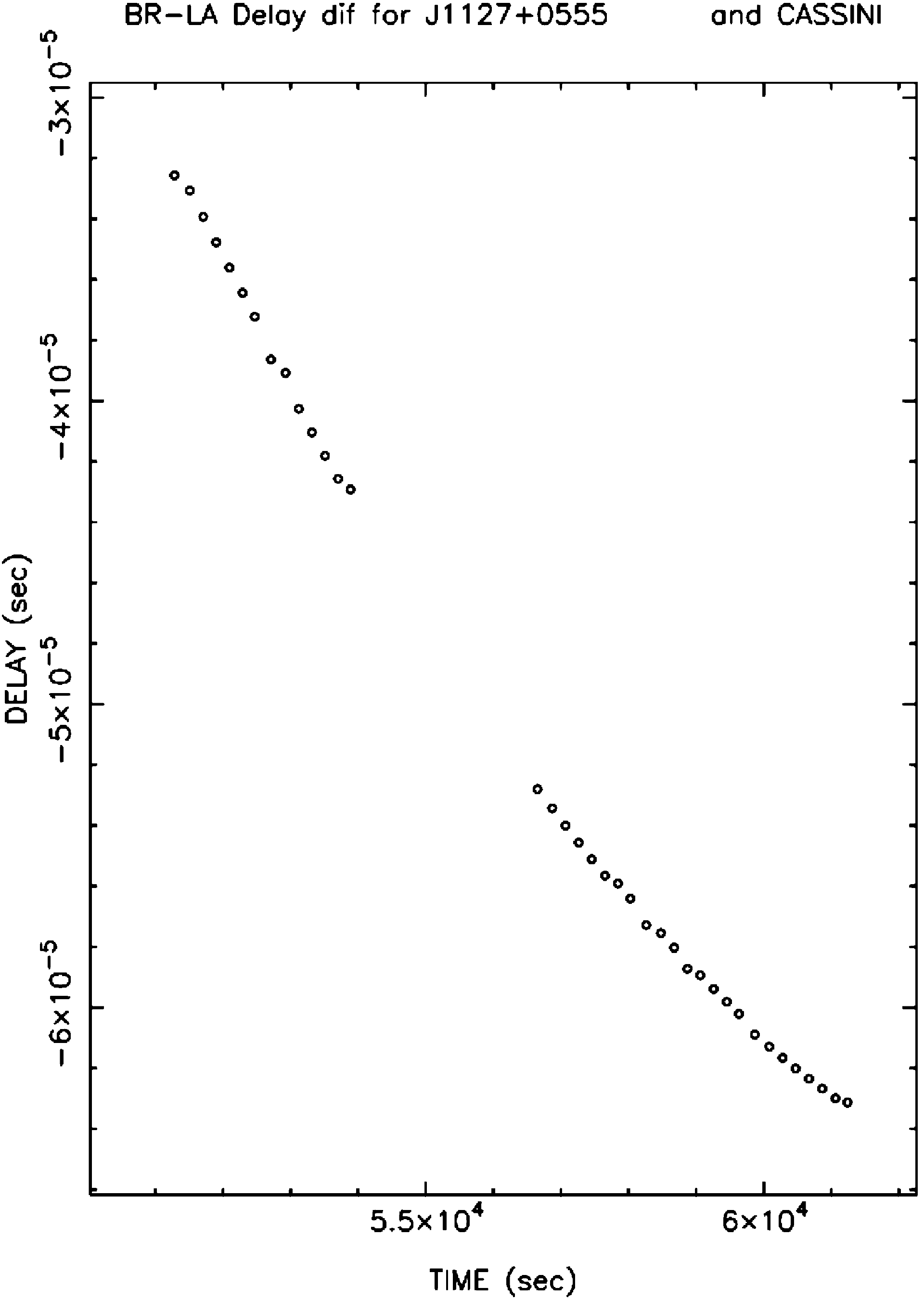}
\caption{Example of total delays on one baselines for 
Cassini and the phase reference source (left), and 
the differential delays between the two sources (right).  The 
horizontal axis spans approximately four hours of time.  \label{fig6}}
\end{figure}

Figure \ref{fig6} shows an example
of the observed total delays for Cassini and the phase reference 
source on one baseline, and the difference in total delays 
between the sources based on interpolation to identical times. 
The variation in differential delay during an observing epoch 
defines the angular offset between Cassini and the phase reference source. 

\section{Results}

Table \ref{tab4} lists our results from phase-referenced imaging of 
Cassini during VLBA observing epochs from October 2006 to 
April 2009.  (Preliminary versions of these results were reported in Jones et al. 
2009a; 2009b.)  The estimated measurement errors, including residual systematic 
errors from troposphere delay calibration and reference sources positions, are $\pm 0.2$ 
mas in right ascension and $\pm 0.4$ mas in declination  The errors are larger in declination 
because of the low declination of Saturn/Cassini, which foreshortens the north-south 
angular resolution of the VLBA.  The two left columns show Cassini image  
position offsets that have been corrected for the one-second time error that 
occurred during our first four epochs.  This correction was made by subtracting 
an angle corresponding to the spacecraft proper motion during one second
of time from each measured coordinate offset.  

There is marginal 
evidence for a small position offset in right ascension (+0.39 $\pm$ 0.12 mas).  
It is important to see if future epochs continue to show this offset.  Given the sometimes 
large uncertainties in the ICRF positions of our phase reference sources, this offset is 
not currently significant.  There is no significant 
evidence for a systematic position offset in declination at this time (+0.47 $\pm$ 0.30 mas).  
The errors quoted here are from the scatter in the measured offsets; they do not 
include the individual offset measurement errors or the reference source position 
errors.  If we consider only the three epochs 
that used J1025+1253, the reference source with the most accurate ICRF position, 
the conclusion is unchanged.  However, additional epochs and improved phase 
reference source positions are needed to verify the reality of any right ascension offset. 

\begin{deluxetable}{cccccc}
\tablewidth{0pt}
\tablecaption{Cassini Position Offsets from Phase-Reference Images \label{tab4}}
\tablehead{
\colhead{Epoch} & \colhead{Observing} & \colhead{Image RA} & 
\colhead{Image DEC} & \colhead{Corrected RA} & \colhead{Corrected DEC} \\
\colhead{Label} & \colhead{Date} & \colhead{Offset} & \colhead{Offset} &
\colhead{Offset} & \colhead{Offset} \\
 & & \colhead{(mas)} & \colhead{(mas)} & \colhead{(mas)} & \colhead{(mas)}}    
\startdata    
BJ061A & 2006 Oct & $+$3.0 & $-$0.4 & $+$0.5 & $+$0.3 \\
BJ061B & 2007 Mar & $-$2.1 & $+$1.3 & $+$0.4 & $+$0.5 \\
BJ061C & 2007 Jun & $+$3.8 & $-$0.4 & $+$0.4 & $+$0.7 \\
BJ061D & 2008 Jan & $-$1.5 & $+$0.6 & $+$0.2 & $-$0.2 \\
BJ061E & 2008 Jun & $+$0.4 & $+$0.0 & $+$0.4 & $+$0.0 \\
BJ061F & 2008 Aug & $+$0.6 & $+$0.0 & $+$0.6 & $+$0.0 \\
BJ061G & 2008 Nov & $+$0.3 & $+$0.9 & $+$0.3 & $+$0.9 \\
BJ061H & 2009 Apr & $+$0.2 & $+$0.8 & $+$0.3 & $+$0.8 \\
\enddata
\end{deluxetable}

Table \ref{tab5} lists the positions of the Cassini spacecraft determined 
from the VLBA total delay measurements, analyzed at JPL.  
This table includes two 
additional epochs from VLBA experiment BR103 in 2004, 
and an epoch in 2009 February from \citet{Fomalont10}.  We 
included data from the 2004 and 2009 February observations 
because they used the same observing technique and instrumentation 
to determine astrometric 
positions for Cassini as the BJ061 experiments used. 

\begin{deluxetable}{cccccc}
\tablewidth{0pt}
\tablecaption{Observed Cassini Positions in ICRF 2.0 Reference Frame \label{tab5}}
\tablehead{
\colhead{Date} & \colhead{Time} & \colhead{Observed} & \colhead{Observed}  \\
\colhead{ } & \colhead{(UTC)} & \colhead{Right Ascension} & \colhead{Declination}}
\startdata
2004\ Sep\ 08 & 18:00:00 & 07$^{\rm h}$45$^{\rm m}$26$^{\rm s}$.89383 & +21$^{\circ}$02\arcmin15.0413\arcsec \\
2004\ Oct\ 20 & 14:00:00 & 07$^{\rm h}$56$^{\rm m}$26$^{\rm s}$.30351 & +20$^{\circ}$38\arcmin57.4360\arcsec  \\
2006\ Oct\ 11 & 17:00:00 & 09$^{\rm h}$39$^{\rm m}$57$^{\rm s}$.36913 & +14$^{\circ}$57\arcmin22.1298\arcsec  \\
2007\ Mar\ 01 & 07:00:00 & 09$^{\rm h}$31$^{\rm m}$30$^{\rm s}$.71957 & +16$^{\circ}$06\arcmin01.4992\arcsec \\
2007\ Jun\ 08 & 00:00:00 & 09$^{\rm h}$31$^{\rm m}$22$^{\rm s}$.39987 & +15$^{\circ}$59\arcmin55.8154\arcsec  \\
2008\ Jan\ 12 & 10:00:00 & 10$^{\rm h}$40$^{\rm m}$54$^{\rm s}$.92717 & +10$^{\circ}$14\arcmin12.4080\arcsec  \\
2008\ Jun\ 14 & 00:00:00 & 10$^{\rm h}$22$^{\rm m}$29$^{\rm s}$.45360 & +12$^{\circ}$00\arcmin31.9588\arcsec  \\
2008\ Aug\ 01 & 22:00:00 & 10$^{\rm h}$40$^{\rm m}$09$^{\rm s}$.03733 & +10$^{\circ}$14\arcmin07.3040\arcsec  \\
2008\ Nov\ 11 & 17:00:00 & 11$^{\rm h}$24$^{\rm m}$12$^{\rm s}$.79125 & +05$^{\circ}$51\arcmin44.0929\arcsec  \\
2009\ Feb\ 11 & 14:00:00 & 11$^{\rm h}$27$^{\rm m}$09$^{\rm s}$.89646 & +05$^{\circ}$58\arcmin57.6093\arcsec  \\
2009\ Apr\ 24 & 06:00:00 & 11$^{\rm h}$09$^{\rm m}$09$^{\rm s}$.77073 & +07$^{\circ}$50\arcmin58.7079\arcsec  \\
\enddata
\tablecomments{Positions are geocentric at the listed observation times.}
\end{deluxetable}

\begin{figure}
\includegraphics[angle=0,scale=0.75]{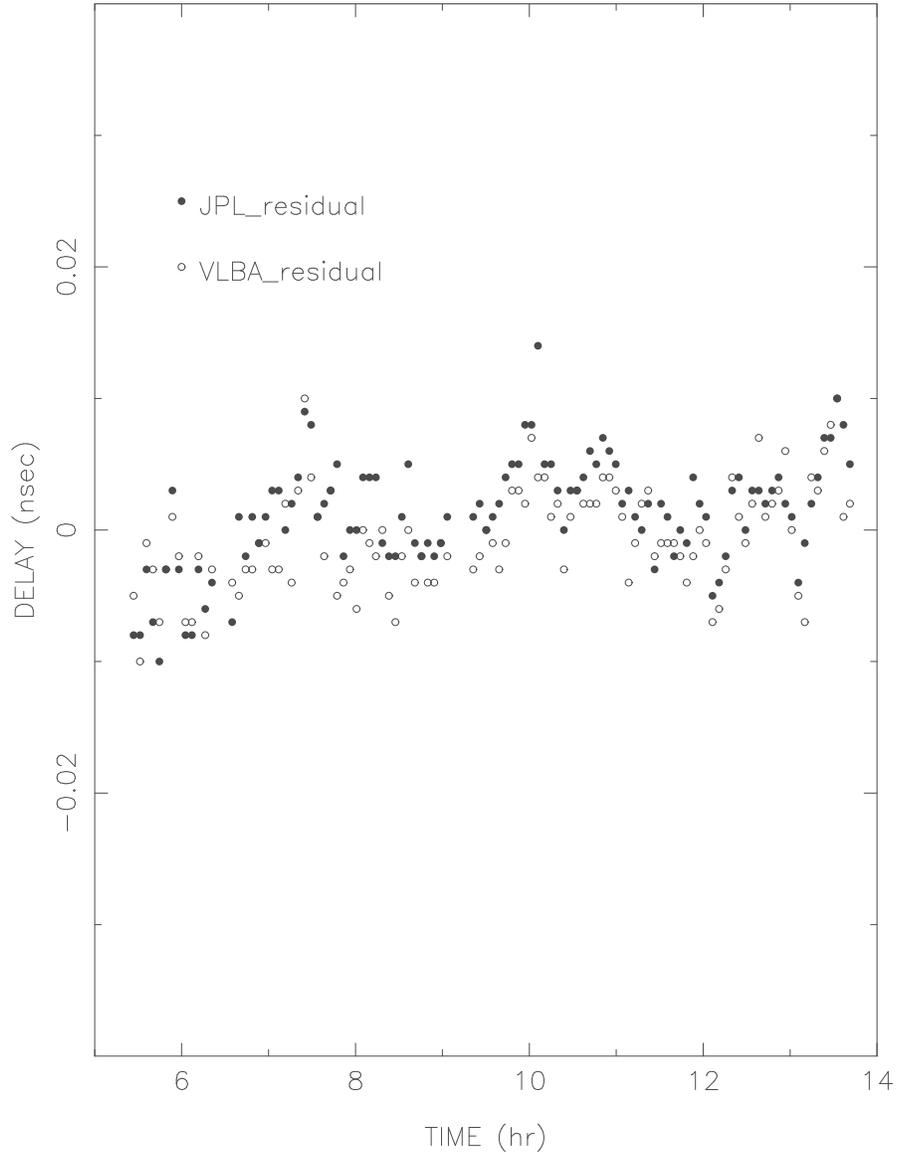}
\caption{Residual delays from the JPL analysis of VLBA total delay 
data and delays determined from phase referenced imaging of Cassini.
\label{fig7}}
\end{figure}

Figure \ref{fig7} shows a comparison of the residual delays on one baseline 
(BR-FD) from one epoch (February 2009), illustrating the level of consistency 
between the two data analysis paths.  The differences are $\sim 2$ ps, which is 
comparable to the numerical precision expected from interpolating the 
planetary ephemeris for Saturn.  (The JPL ephemeris is integrated in 
quadruple precision, but the results are stored in double precision so the 
numerical noise in the $\sim$5000 light second distance to Saturn is a few ps.)
The overall slope from about -5 ps to +5 ps during this epoch corresponds 
to the residual Cassini position offset from the {\it a priori} position, and is 
clearly seen equally in the delays from both analyses.  

\clearpage

The derived J2000 (ICRF 2.0) positions of the Saturn system barycenter 
from VLBA observations of Cassini, including detailed  
Cassini orbit reconstructions, are listed in Table \ref{tab6}.  
This table is the primary result of our observations.  
The Cassini orbital trajectory is produced by numerical integration 
of the equations of motion as part of a global Saturn ephemeris and 
gravity field solution (e.g., \citet{Antreasian06}; \citet{Jacobson06}).  These solutions 
include a large number of historical and recent observations.  The equations of 
motion for Cassini include Newtonian accelerations due to the Sun, planets, and 
the Saturnian satellites, relativistic perturbations due to the Sun, Jupiter, and 
Saturn, and perturbations due to the oblateness of Saturn.  Non-gravitational 
forces such as attitude control and trajectory correction maneuvers and 
solar radiation pressure are also modeled.  Saturn's satellite dynamics include 
the mutual interactions of the satellites and perturbations due to the Sun, 
Jupiter, Uranus, Neptune, and Saturn's oblateness.  

\begin{deluxetable}{cccccc}
\tablewidth{0pt}
\tablecaption{Observed Saturn Barycenter Positions in ICRF 2.0 Reference Frame \label{tab6}}
\tablehead{
\colhead{Date} & \colhead{Time} & \colhead{Observed} & \colhead{Observed} &
\colhead{Error in} & \colhead{Error in} \\
\colhead{ } & \colhead{(UTC)} & \colhead{Right Ascension} & 
\colhead{Declination} & \colhead{R.A. (s)} & \colhead{Dec. (\arcsec)}}
\startdata
2004\ Sep\ 08 & 18:00:00 & 07$^{\rm h}$43$^{\rm m}$57$^{\rm s}$.853974 & +21$^{\circ}$06\arcmin11.47431\arcsec & 0.000073 & 0.00052 \\
2004\ Oct\ 20 & 14:00:00 & 07$^{\rm h}$55$^{\rm m}$52$^{\rm s}$.671888 & +20$^{\circ}$38\arcmin20.56188\arcsec & 0.000006 & 0.00020 \\
2006\ Oct\ 11 & 17:00:00 & 09$^{\rm h}$39$^{\rm m}$54$^{\rm s}$.457150 & +14$^{\circ}$57\arcmin55.39349\arcsec & 0.000033 & 0.00113 \\
2007\ Mar 1  & 07:00:00  & 09$^{\rm h}$31$^{\rm m}$40$^{\rm s}$.709327 & +16$^{\circ}$02\arcmin49.54177\arcsec & 0.000029 & 0.00091 \\
2007\ Jun\ 08 & 00:00:00 & 09$^{\rm h}$31$^{\rm m}$40$^{\rm s}$.531553 & +15$^{\circ}$59\arcmin06.93823\arcsec & 0.000029 & 0.00094 \\
2008\ Jan\ 12 & 10:00:00 & 10$^{\rm h}$41$^{\rm m}$00$^{\rm s}$.869116 & +10$^{\circ}$11\arcmin45.98652\arcsec & 0.000010 & 0.00031 \\
2008\ Jun\ 14 & 00:00:00 & 10$^{\rm h}$22$^{\rm m}$29$^{\rm s}$.258227 & +11$^{\circ}$59\arcmin01.78134\arcsec & 0.000007 & 0.00028 \\
2008\ Aug\ 01 & 22:00:00 & 10$^{\rm h}$40$^{\rm m}$07$^{\rm s}$.840686 & +10$^{\circ}$12\arcmin49.68886\arcsec & 0.000009 & 0.00022 \\
2008\ Nov\ 11 & 17:00:00 & 11$^{\rm h}$24$^{\rm m}$07$^{\rm s}$.553645 & +05$^{\circ}$51\arcmin34.99358\arcsec & 0.000065 & 0.00173 \\
2009\ Feb\ 11 & 14:00:00 & 11$^{\rm h}$27$^{\rm m}$15$^{\rm s}$.292877 & +05$^{\circ}$56\arcmin37.40025\arcsec & 0.000016 & 0.00051 \\
2009\ Apr\ 24 & 06:00:00 & 11$^{\rm h}$09$^{\rm m}$02$^{\rm s}$.825609 & +07$^{\circ}$52\arcmin58.01128\arcsec & 0.000017 & 0.00053 \\
\enddata
\tablecomments{Positions are geocentric at the listed signal reception times.  These  
define the direction vector from the Earth geocenter at signal reception time to Saturn's  
position at signal transmission time (earlier than signal reception by the light travel time 
from Saturn).  Thus, no aberration or relativistic light deflection has been applied.}
\end{deluxetable}

Table \ref{tab7} gives the post-fit residuals for each epoch in Table \ref{tab6} 
with respect to the DE422 ephemeris, which was fit to the VLBA data.  For 
comparison, residuals are also give with respect to the widely used DE405 
ephemeris.  

\begin{deluxetable}{cccccc}
\tablewidth{0pt}
\tablecaption{VLBA Position Residuals for Saturn Barycenter \label{tab7}}
\tablehead{
\colhead{Date} & \colhead{Time} & \colhead{$\alpha$-DE422} & 
\colhead{$\delta$-DE422} & \colhead{$\alpha$-DE405} & \colhead{$\delta$-DE405} \\
 & \colhead{(TDB)} & \colhead{(\arcsec)} & \colhead{(\arcsec)} & \colhead{(\arcsec)} & \colhead{(\arcsec)}}
\startdata
2004\ Sep\ 08 & 18:00:00 &  -0.00181 & -0.00029  & 0.11810 & -0.04001 \\
2004\ Oct\ 20 & 14:00:00 &  -0.00004  & 0.00012  & 0.12933 & -0.04415 \\
2006\ Oct\ 11 & 17:00:00 &   0.00013  & -0.00007  & 0.13055 & -0.04509 \\
2007\ Mar\ 1  & 07:00:00 &  0.00010  & 0.00002  & 0.15703  &  -0.04761 \\
2007\ Jun\ 08 & 00:00:00 &   0.00024 &  0.00008  & 0.13570 & -0.03817 \\
2008\ Jan\ 12 & 10:00:00 &   0.00005  & -0.00001  & 0.14914 & -0.04794 \\
2008\ Jun\ 14 & 00:00:00 &   0.00007 & -0.00005  & 0.13594 & -0.03678 \\
2008\ Aug\ 01 & 22:00:00 &  0.00002 & -0.00010  & 0.12708 & -0.03553 \\
2008\ Nov\ 11 & 17:00:00 &   0.00005 &  0.00080 &  0.12868 & -0.03666 \\
2009\ Feb\ 11 & 14:00:00 &  -0.00006 &  0.00024  & 0.14930 & -0.04090 \\
2009\ Apr\ 24 & 06:00:00 &  -0.00005 &  0.00045 &  0.14785 & -0.03653 \\
\enddata
\tablecomments{TDB = Barycentric Dynamic Time, a general relativistic 
coordinate time centered on the solar system barycenter.}
\end{deluxetable}

\begin{figure}
\includegraphics[angle=0,scale=1.00]{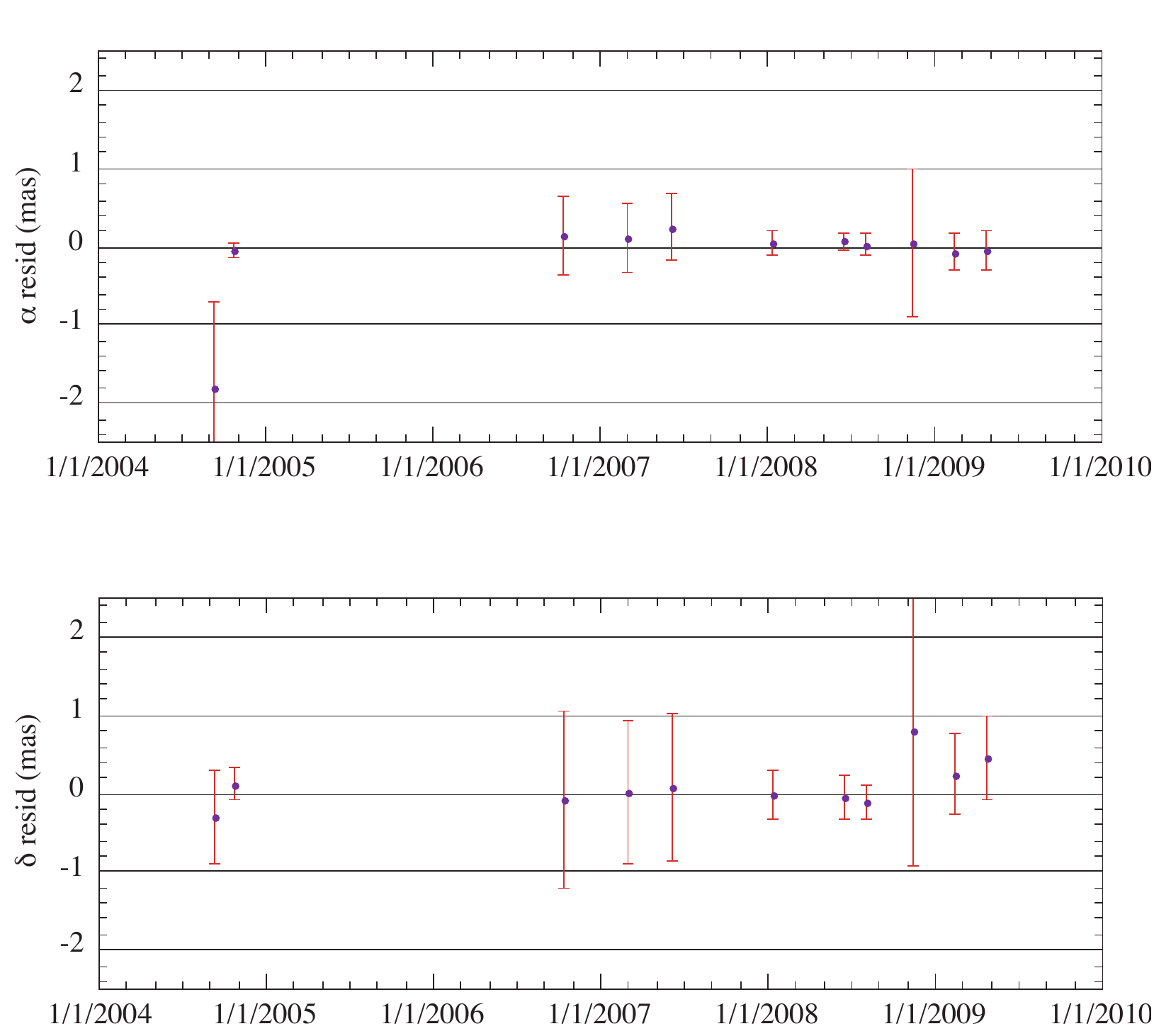}
\caption{Saturn positions based on VLBA data compared to the DE422 
JPL Saturn ephemeris positions.  The error estimates include expected uncertainties in the phase 
reference source positions prior to our improved tie of these sources 
to the ICRF, and also include errors in the solutions for the Cassini 
orbit about Saturn.  The first two data points were obtained by a separate observing 
program; the first data point was obtained prior to Cassini orbit insertion  
at Saturn.
\label{fig8}}
\end{figure}

The observed positions of the Saturn barycenter with 
respect to the DE422 planetary ephemeris are shown in Figure \ref{fig8}.   
The DE422 ephemeris is the result of fitting the DE421 ephemeris 
(\citet{Folkner08}; \citet{Folkner09}) to the new Cassini VLBA data, along with 
about 18 months of additional tracking data from Venus Express, Mars Reconnaissance 
Orbiter, Mars Express, Mars Odyssey, and CCD observations of the outer planets.  
Errors are estimated 
independently by the ephemeris fitting program used at JPL, and include 
uncertainties in both the VLBA measurements and the position of Cassini 
with respect to the Saturn barycenter.  The weighted mean 
offset after fitting is less than 0.2 mas in right ascension and 0.3 mas 
in declination, consistent with the expected uncertainties.  

\clearpage

\section{Conclusions}

We have demonstrated repeatable phase referenced astrometry of 
the Cassini spacecraft using the VLBA, and verified consistent position 
determinations from direct imaging and from total delay measurements. 
Future observations will increase the time span of accurate position 
measurements, leading to ever improving constraints on the planetary 
ephemeris.  In addition, continuing improvement in the accuracy of 
phase reference source positions will allow a more accurate tie of 
our Cassini positions to the ICRF.  

The Cassini mission has recently been extended until 2017, with further 
extensions likely in the future.  By extending our VLBI observations 
beyond 2012 we will have high accuracy measurements over more 
than a quarter of Saturn's orbital period.  The error in determining the 
plane of Saturn's orbit (latitude) decreases rapidly as the time span 
of observations approaches 1/4 of the orbital period.  The error in 
longitude decreases approximately linearly with time span.  

The next mission to the outer planets will be the JUNO mission to Jupiter. 
This orbiting mission will provide an opportunity to use the same phase 
referenced astrometry techniques with the VLBA, and thereby improve 
the ephemeris of Jupiter in a similar manner.   

\acknowledgments

We are grateful to Larry Teitelbaum for support of this project 
through the Advanced Tracking and Observational Techniques office of JPL's 
Interplanetary Network Directorate, and to John Benson and the VLBA 
operations staff at NRAO for their excellent support of these observations.  
We also thank Peter Antreasian and Fred Pelletier at JPL for providing the 
reconstructed Cassini orbit files used for data correlation at NRAO.    
The anonymous referee's comments led to significant improvements 
in the paper.
The VLBA is a facility of the National Radio Astronomy Observatory, which 
is operated by Associated Universities, Inc., under a cooperative agreement 
with the National Science Foundation.  Part of this research was carried out 
at the Jet Propulsion Laboratory, California Institute of Technology, under 
contract with the National Aeronautics and Space Administration.  

{\it Facilities:} \facility{VLBA}.

\clearpage

\end{document}